\documentclass[english,two column, showpacs]{revtex4}
\usepackage[T1]{fontenc}
\usepackage[latin1]{inputenc}
\usepackage{graphicx}
\usepackage{amsmath}
\usepackage{amssymb}
\usepackage{amsfonts}
\def\a{\alpha}

\def\e{\varepsilon}

\def\r{\rho}

\def\be{\begin{equation}}
\def\ee{\end{equation}}
\def\bea{\begin{eqnarray}}
\def\eea{\end{eqnarray}}
\def\nn{\nonumber}
\def\lb{\label}

\begin{document}

\title{Superconducting junctions from non-superconducting doped CuO$_2$ layers}
\author{V. M. Loktev$^1$ and Yu. G. Pogorelov$^2$}
\affiliation{$^1$N.N. Bogolyubov Institute for Theoretical Physics, Metrologichna 14b, Kiev
03134, Ukraine,\\ $^2$IFIMUP, Universidade do Porto, R. Campo Alegre, 687, Porto 4169-007}

\begin{abstract}
The theoretical approach proposed recently for description of redistribution of  electronic
charge in multilayered selectively doped systems is modified for a system with finite number
of layers. A special attention is payed to the case of a finite heterostructure made of
copper-oxide layers which are all non-superconducting (including non-conducting) because of
doping levels being beyond the well-known characteristic interval for superconductivity.
Specific finite structures and doping configurations are proposed to obtain atomically thin
superconducting heterojunctions of different compositions.
\end{abstract}

\pacs{74.72.-h, 74.78.Fk, 74.25.Jb, 74.45.+c}
\maketitle\

An interesting area in nanoengineering of materials was opened in a series of experiments
by Bozovi\v{c} \emph{et al} \cite{Boz1} on atomically perfect stacks of selectively doped
perovskite layers. These and some other papers \cite{Goz1,Boz2} mainly used periodic
multilayered structures where essential new electronic effects, as interface SC between
nominally non-SC layers \cite{Boz2}, appeared due to charge redistribution between layers
and related shifts of in-plane energy bands. The basic condition for SC to appear within
few perovskite layers or even in a single layer is that the local density of hole charge
carriers occurs within a definite, rather narrow, interval: $p_{min} \geq p \geq p_{max}$
with $p_{min} \approx 0.07$ and $p_{max} \approx 0.2$ (carriers per site). The required
density distribution results from the corresponding shifts of in-plane energy bands by local
Coulomb potentials. A simple theoretical model for such processes was proposed \cite{lok},
combining a discrete version of Poisson equation for potential with a band-structure
modified self-consistent Thomas-Fermi charge density. This approach gives exact solutions
for infinite periodical and some other unbounded systems. However recent studies
\cite{Goz2, Smad, Log} showed that pronounced modification of electronic ground state and
related SC transitions can be obtained either in stacks of finite (and small) number of
layers which is quite promising for practical applications in nanoengineered composite
devices. The following consideration aims on an extension of the previous model on an
arbitrary layered system and establishing criteria for its optimum SC performance. This
line of study can be seen as a practical realization of long envisaged Ginzburg's program
for ultrathin superconducting states \cite{ginz}.

Following the same assumptions as in Ref. \cite{lok}, we express local charge density in
\emph{j}-th layer: $\r_j = e\left(p_j - x_j\right)$, through the densities $p_j$ of mobile
holes and $x_j$ of ionized dopants ($e$ is the elementary charge) and then present the
potential difference $\varphi_{j+1} - \varphi_j$ between neighbor layers as:
\be
 \varphi_{j+1} - \varphi_j = \frac{2\pi c}{\e_{eff}a^2} \left(\sum_{l=j+1}^N \r_l -
  \sum_{l=1}^j \r_l\right).
   \lb{eq1}
    \ee
This value is obtained considering the electric field $E_{j,j+1}$ in the $j,j+1$ spacer as
the geometric sum of fields $E_l$ emitted by each \emph{l}-th charged layer: $E_l = 2\pi
\r_l/\left(\e_{\rm eff}a^2\right)$, on both sides of this spacer. Eq. \ref{eq1} involves
the in-plane and \emph{c}-axis lattice constants $a$ and $c$, and $\e_{\rm eff}$ is the
(static) dielectric constant that effectively reduces the Coulomb field in the
\emph{c}-direction. Eq. \ref{eq1} would be exact for a stack of mathematical planes, with
uniform in-plane charge densities $\r_j$ and separation $c$, and it should be a reasonable
model for real ${\rm La}_{2-x}{\rm Sr}_x{\rm CuO}_4$ layers where $p_j$ delocalized holes
and $x_j$ localized dopants are distributed in different atomic planes within the period
$c$ of \emph{j}th layer. The adopted form of purely dielectric screening is justified in
neglect of \emph{c}-hopping processes, accordingly to their above mentioned weakness, also
this model neglects polarization effects from the insulating substrate. We note that the
charge densities $\r_j$ naturally vanish in uniformly doped ($p_j = x_j$), including
undoped ($p_j = x_j = 0$), systems.
\begin{figure}
\center\includegraphics[width=8cm]{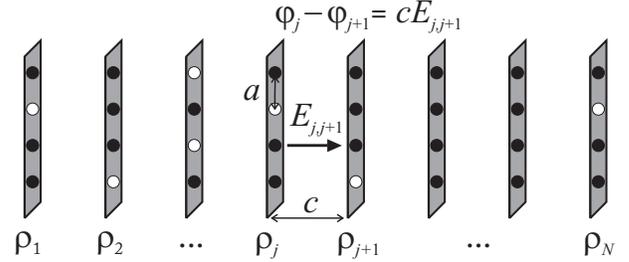}\\
\caption{Schematic of nanostructured system with selectively introduced dopants (white
circles) into each of $N$ layers of ${\rm La}_2{\rm CuO}_4$.}
  \label{fig1}
\end{figure}
Otherwise, the hole carrier density $p_j$ can be related to the local potential $\varphi_j$
using the respective density of states (DOS) $g_j(\e)$:
\be
 p_j = 2\int_{\e_{\rm F}}^{W/2 - e\varphi_j}g_j(\e)d\e
  \lb{eq2}
   \ee
with the spin factor 2 (this zero-temperature formula is justified for all the considered
temperatures $T \lesssim T_c$). Thus the role of \emph{c}-hopping in this model is reduced
to establishing the common Fermi level $\e_{\rm F}$ for all the layers. Using the simplest
approximation of rectangular DOS: $g_j(\e) = 1/W$ within the bandwidth $W$, we arrive at
the linear relation between $p_j$ and $\varphi_j$:
\be
 e\varphi_j = \frac{1-p_j} 2 W  - \e_{\rm F}.
  \lb{eq3}
   \ee
Then, inserting Eq. \ref{eq3} into Eq. \ref{eq1}, we obtain the linear relation between
carrier and dopant densities, referred to the $j,j+1$ spacer:
\be
 p_{j+1} - p_j = \frac \a 2\left[\sum_{l = 1}^j \left(p_l - x_l\right) -
 \sum_{l=j+1}^N \left(p_l - x_l\right)\right],
   \lb{eq4}
    \ee
where the dimensionless quantity:
\be
 \a = \frac{8\pi c e^2}{W\e_{\rm eff}a^2}
  \lb{eq5}
   \ee
is the single material parameter of the model. Finally, subtracting the relations, Eq.
\ref{eq4}, for $j,j+1$ and $j - 1,j$ spacers leads to simple linear equations for the hole
carrier densities in neighbor layers only:
\be
 p_{j+1} + p_{j-1} - (2 + \a)p_j = -\a x_j.
  \lb{eq5}
   \ee
The advantage of Eq. \ref{eq5} against possible analogous relations between the potentials
$\varphi_j$ is in avoiding the need to know the Fermi level position (doping dependent).
For an infinite stack of layers, summing these equations in all $j$ would automatically
assure the total electroneutrality condition $\sum_j \r_j = 0$, and this was just the way
used in Ref. \cite{lok} to obtain a more detailed alternative to the phenomenological
Thomas-Fermi treatment \cite{Smad}.  However, for a finite stack of $j = 1,\dots,N$ layers,
this condition should be additionally imposed besides the $N - 1$ relations, Eq. \ref{eq4},
in order to completely define all the $N$ densities $p_j$. Since Eq. \ref{eq5} in this case
only applies for the \emph{internal} layers $j = 2,\dots,N-1$, one needs two more equations
which can be obtained from Eq. \ref{eq4} for terminal layers, $j = 1$ and $j = N -1$, under
the electroneutrality condition:
\bea
 (1 + \a)p_1 - p_2 & = & \a x_1\nn \\
  - p_{N -1} + (1 + \a)p_N & = & \a x_N.
  \lb{eq6}
 \eea
Thus the non-uniform linear system of $N$ Eqs. \ref{eq5}, \ref{eq6} can be presented in the
matrix form as:
\be
 \left(1 + \a^{-1}\hat L\right)\overrightarrow{p}  = \overrightarrow{x},
  \lb{eq7}
   \ee
where the finite $N$-stack of layers generates the "Laplacian" matrix:
\be
 \hat L = \left(\begin{array}{cccccc}
                    1 & -1 & 0 & 0 & \dots & 0 \\
                    -1 & 2 & -1 & 0 & \dots & 0 \\
                     0 & -1 & 2 & -1 & \dots & 0 \\
    \dots & \dots & \dots & \dots & \dots & \dots \\
                       0 & \dots & 0 & -1 & 2 & -1 \\
        0 & \dots & 0 & 0 & -1 & 1\end{array}\right)
        \lb{eq8}
         \ee
and the $N$-vectors:
\[\overrightarrow{p} = \left(\begin{array}{c}
        p_1 \\
        \dots \\
        p_N
      \end{array}\right),\qquad{\rm and}\qquad \overrightarrow{x}=\left(\begin{array}{c}
        x_1 \\
        \dots \\
        x_N
      \end{array}\right).\]
\begin{figure}
\center\includegraphics[width=7cm]{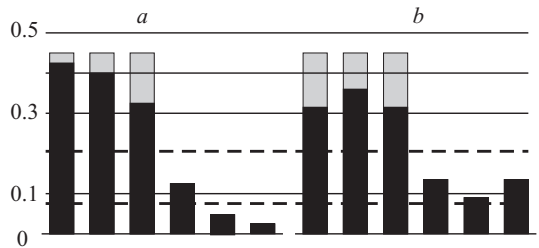}\\
\caption{Modulated electronic configurations by shifted energy bands for the samples of
selectively doped layered systems. a) For a finite stack of $N = 6$ layers with doping
levels $x_1 = x_2 = x_3 = 0.45$, $x_4 = x_5 = x_6 = 0$ (light columns) and the localization
parameter $\a = 1$, the carrier densities (dark columns) are calculated from Eqs. \ref{eq7},
\ref{eq8}. b) For an unlimited system with periodic repetition of the same stack, those
are calculated from Eqs. \ref{eq7}, \ref{eq9}. The dashed lines mark the values $p_{min}$
and $p_{max}$, delimiting the interval of carrier densities where superconductivity should
exist.}
  \label{fig2}
\end{figure}
It is seen from Eq. \ref{eq7} that $\a$ plays the role of localization parameter: the carrier
density strictly coincides with the doping distribution in the limit $\a \to \infty$, otherwise
it is spread beyond this distribution, the stronger the smaller $\a$ is. Generally, the standard
solution $\overrightarrow{p} = \hat R(\a)\overrightarrow{x}$ with the resolvent $\hat R =
\left(1 + \a^{-1}\hat L\right)^{-1}$ gives the densities $p_j$ in terms of the doping levels
$x_j$ and of the localization parameter $\a$ as for the instance in Fig. \ref{fig2} with $N =
6$, $\a = 1$ (a reliable estimate for real La$_2$CuO$_4$ \cite{pog,kast, chen}) and $x_1 = x_2
= x_3 = 0.45,\, x_4 = x_5 = x_6 = 0$. It is of interest to compare this solution to that for
an unlimited system with the same distribution of dopants $x_j$ but periodically repeated, so
that the $\hat L$ matrix is replaced by its periodic version:
\be
 \hat L' = \left(\begin{array}{cccccc}
                    2 & -1 & 0 & \dots  & 0 & -1 \\
                    -1 & 2 & -1 & 0 & \dots & 0 \\
                     0 & -1 & 2 & -1 & \dots & 0 \\
    \dots & \dots & \dots & \dots & \dots & \dots \\
                       0 & \dots & 0 & -1 & 2 & -1 \\
        -1 & 0 & \dots & 0 & -1 & 2\end{array}\right).
        \lb{eq9}
         \ee
A notable difference in the resulting distributions of densities $p_j$ is seen in Fig.
\ref{fig2}a,b. An important feature of superconducting layers formed in each of these
structures (4th layer in Fig. \ref{fig2}a and 4th and 6th layers in Fig. \ref{fig2}b)
is that they are realized in nominally undoped La$_2$CuO$_4$ and thus can be expected
almost free of undesirable disorder effects (as scattering by defects and consequent
fluctuations of the SC order parameter \cite{lang}).

\begin{figure}
\center\includegraphics[width=8cm]{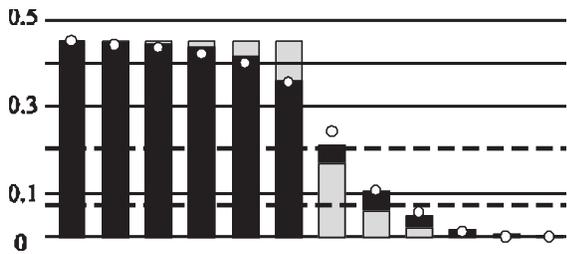}\\
\caption{Comparison of calculated (dark columns) and experimentally determined (white
circles) carrier densities for the real system with $N = 12$, the doping levels $x_1 = x_2
= x_4 = x_5 = x_6 = 0.45,\, x_7 = 0.17,\, x_8 = 0.06, x_9 = 0.02,\,x_{10} = x_{11} = x_{12}
= 0$ (light columns) and the localization parameter $\a = 1$.}
  \label{fig3}
\end{figure}

Also a comparison of such calculation with the experiment data \cite{Log} is presented in
Fig. \ref{fig3}, using the real values of structure parameters: $N = 12$ and $x_1 = x_2 =
x_4 = x_5 = x_6 = 0.45,\, x_7 = 0.17,\, x_8 = 0.06, x_9 = 0.02$ (the three last values are
due to the interdiffusion of Sr dopants into nominally undoped La$_2$CuO$_4$ layers), $x_{10}
= x_{11} = x_{12} = 0$ and the same localization parameter $\a = 1$ as before. The resulting
distribution of carrier densities (black columns) displays an excellent coincidence with its
measured values. In particular, the carrier density occurs within the SC interval just in the
$N = 8$ layer with the value $p_8 \approx 0.106$ that just corresponds to the observed transition
temperature $T_c \approx 32$ K when used in the phenomenological formula \cite{lok} $T_c(p) =
\left(p - p_{min}\right)\left(p_{max} - p\right)T^\ast$ with $T^\ast \approx 9000$ K.

From such a good agreement, one expects that this approach can be also used for design of
new configurations with tailored superconducting performance. In particular, an interesting
possibility is to build ultra-quantum heterojunctions of two types: superconductor-
insulator-superconductor (SIS) and superconductor-normal metal-superconductor (SNS), each
component being restricted to a single cuprate layer. Such junctions could realize a single
layer limit of already discussed thicker SIS and SNS nanostructures with giant proximity effect
\cite{Boz4}. It should be noted that since the localization parameter $\a$ value (see Eq.
\ref{eq5}) is rather fixed by the choice of the building material (with $\a \approx 1$ for
La$_2$CuO$_4$), the practical control parameters in this process must be the total number
$N$ of layers in the stack and particular doping levels $x_j$ in each layer.

Thus, one possible simple structure to produce a SNS junction can consist of $N = 5$ cuprate
layers with the doping levels defined for instance as: $x_1 = x_5 = 0.45,\, x_2 = x_3 = x_4
= 0$ (that is, nominally all non-superconducting). From Eq. \ref{eq7}, the resulting carrier
density distribution: $p_1 = p_5 \approx 0.29$, correspond to nomal (overdoped) metal layers,
$p_2 = p_4 \approx 0.12$ to superconducting layers with a high enough transition temperature
$T_c^{high} \approx 37$ K, separated by the layer with $p_3 \approx 0.08$ and low transition
temperature $T_c^{low} \approx 11$ K. Then, in the temperature range $T_c^{low} < T <
T_c^{high}$ one should obtain a SNS heterostructure. If so, the quasiparticle spectrum of
this junction will present a peculiar combination of gapped (a kind of size quantization)
and gapless branches with interesting IR absorption and electric transport properties.

As to the SIS heterostructure, it is rather difficult to be obtained in the suggested $N =
5$ stack, but it can be achieved, e.g., by adding one more undoped layer (i.e., from 3 to
4) to the above structure, or introducing \emph{electronic} doping in the central layer
(making $x_3 < 0$). Unlike the above SNS case, the resulting SIS junction should display a
quasiparticle spectrum with gapped branches only.

In conclusion, an extension of the recent electrostatic model for charge redistribution in
non-uniformly doped multilayered systems is proposed for finite (mostly small) number of
layers. Formal solutions of this model are mainly analyzed in the parameter range actual
for the experimentally investigated ${\rm La}_{2-x}{\rm Sr}_x{\rm CuO}_4$ multilayers. The
distinctions of finite stacks from previously studied unlimited or cyclic systems are
indicated. Some new specific arrangements of doped and undoped layers are suggested for
realization of artificial atomically thin heterostructures with unusual electronic excitation
spectrum, potentially interesting for applications in nanocomputing devices. Such artificial
structures may present also an interest for their behavior under applied magnetic field.

This work was partially supported by the Special Program of Fundamental Research of the
Department of Physics and Astronomy of NAS of Ukraine. The authors are grateful to
I. Bozovi\v{c} for reading the manuscript and valuable discussion.

\end{document}